%% file: main.tex
\documentclass[a4paper, amsfonts, amssymb, amsmath, reprint, showkeys, nofootinbib, twoside]{revtex4-1}
\usepackage[english]{babel}
\usepackage[utf8]{inputenc}
\usepackage{here}
\usepackage{lineno}
\usepackage[colorinlistoftodos, color=green!40, prependcaption]{todonotes}
\usepackage{comment}
\input{preamble}
\begin{document}
\title{Angular distribution of $\gamma$ rays from the p-wave resonance of $^{118}$Sn}

\author{J.~Koga$^1$}
    \email[Correspondence email address: ]{junkoga@kyudai.jp}
\author{S.~Takada$^1$}
\author{S.~Endo$^{2,3}$}
\author{H.~Fujioka$^4$}
\author{K.~Hirota$^3$}
%    \email[Present address: ]{High Energy Accelerator Research \\Organization, 1-1 Oho, Tsukuba, Ibaraki 305-0801, Japan}
\author{K.~Ishizaki$^3$}
\author{A.~Kimura$^2$}
\author{M.~Kitaguchi$^3$}
\author{Y.~Niinomi$^3$}
\author{T.~Okudaira$^3$}
\author{K.~Sakai$^2$}
\author{T.~Shima$^5$}
\author{H.~M.~Shimizu$^3$}
\author{Y.~Tani$^4$}
\author{T.~Yamamoto$^3$}
\author{H.~Yoshikawa$^5$}
\author{T.~Yoshioka$^1$}
    \email[Correspondence email address: ]{yoshioka@phys.kyushu-u.ac.jp}

    \affiliation{$^1$Kyushu University, 744 Motooka, Nishi-ku, Fukuoka 819-0395, Japan}
    \affiliation{$^2$Japan Atomic Energy Agency, 2-1 Shirane, Tokai 319-1195, Japan}
    \affiliation{$^3$Nagoya University, Furocho, Chikusa, Nagoya 464-8062, Japan}
    \affiliation{$^4$Tokyo Institute of Techonology, Meguro, Tokyo 152-8511, Japan}        
    \affiliation{$^5$Osaka University, Ibaraki, Osaka 567-0047, Japan}

\date{\today} % Leave empty to omit a date

\begin{abstract}
%\lipsum[1]
The neutron energy-dependent angular distribution of $\gamma$ rays from $^{117}{\rm Sn}(n,\gamma)$ reaction was measured with germanium detectors and a pulsed neutron beam.
The angular distribution was clearly observed in $\gamma$-ray emissions with an energy of 9327~keV which corresponds to the transition from a neutron resonance of $^{117}{\rm Sn}+n$ to the ground state of $^{118}{\rm Sn}$.
The angular distribution causes an angular-dependent asymmetric resonance shape.
An asymmetry $A_{\rm LH}$ was defined as $(N_{\rm L}-N_{\rm H})/(N_{\rm L}+N_{\rm H})$, where $N_{\rm L}$ and $N_{\rm H}$ are integrated values for lower- and higher-energy regions of a neutron resonance, respectively.
%We found that the $A_{\rm LH}$ has the angular dependence of $(A\cos\theta + B)$, where $\theta$ is the $\gamma$-ray emission angle, with $A=0.47 \pm 0.05({\rm stat.}) \pm 0.03({\rm sys.})$ in the 1.33~eV p-wave resonance.
We found that the $A_{\rm LH}$ has the angular dependence of $(A \cos \theta_\gamma +B)$, where $\theta_\gamma$ is the $\gamma$-ray emission angle with respect to the incident neutron momentum, with $A=0.394 \pm 0.073$ and $B = 0.118 \pm 0.029$ in the 1.33~eV p-wave resonance. 
%Additionally, we use the result to evaluate the possible enhancement of the violation of time-reversal symmetry in the vicinity of the p-wave resonance.
\end{abstract}

%\keywords{first keyword, second keyword, third keyword}

\maketitle
%\linenumbers
\input{sections/section01.tex}  %I believe leaving the sections in separate files is more organized, change it if you desire 
\input{sections/section02.tex}
\input{sections/section03.tex}

\input{sections/section04.tex}
\input{sections/section05.tex}
\input{sections/acknowledgements.tex}

\input{sections/Reference}
%\appendix*
%\input{sections/appendix1.tex}

\end{document}

%% file: preamble.tex
\usepackage{amsthm}
\usepackage{mathtools}
\usepackage{physics}
\usepackage{xcolor}
\usepackage{graphicx}
\usepackage[left=23mm,right=13mm,top=35mm,columnsep=15pt]{geometry} 
\usepackage{adjustbox}
\usepackage{placeins}
\usepackage[T1]{fontenc}
\usepackage{lipsum}
\usepackage{csquotes}

%% file: sections/section01.tex
\section{Introduction} \label{sec:introduction} 

Neutron resonances in medium-mass nuclei are expected as one of the sensitive probes for symmetry-breaking physics.
A large parity violation (P-violation) has been observed as the helicity dependence of the neutron-capture cross section at p-wave resonances in various nuclei~\cite{PhysRep.354.157}.
These large P-violating effects are explained as a result of the interference between the amplitudes of p-wave resonances and the neighboring s-wave resonances ($sp$-mixing model).
In addition, it is predicted that the violation of time-reversal symmetry (T-violation) can be also enhanced by the same mechanism based on the $sp$-mixing model~\cite{PhysRep.212.77}.
V.~V.~Flambaum {\it et al.} reported that the $sp$-mixing causes a neutron energy-dependent angular distribution of $\gamma$ rays in the vicinity of a p-wave resonance with respect to the incident neutron momentum~\cite{NuclPhysA.435.352}.

The first measurement of angular distribution was conducted for the $(n,\gamma)$ reactions of $^{117}{\rm Sn}+n$ at the IBR-30 reactor, located at the Laboratory of Neutron Physics, JINR in 1985~\cite{JINR.17.22}.
The angular distribution was observed using two NaI(Tl) spectrometers at $90+\theta$ and $90 - \theta$ degrees with respect to the neutron beam axis.
Because an energy difference between the ground and the first excited states of $^{117}{\rm Sn}+n$ is more than 1000~keV, even the $\gamma$-ray energy resolution of NaI spectrometers can separate their states.
In addition, a left-right asymmetry in the emission of $\gamma$ rays was obtained by the measurement with polarized neutron beam~\cite{JINR.17.22}.
%In addition, other angular distribution was obtained by the measurement with polarized neutron beam~\cite{JINR.17.22}.

Recently, thanks to a high $\gamma$-ray energy resolution of germanium detectors at the Accurate Neutron-Nucleus Reaction and measurement (ANNRI) in the Japan Proton Accelerator Research Complex (J-PARC), we can measure the angular distribution for each transition with a large signal to background ratio~\cite{PhysRevC.97.034662, PhysRevC.101.064624, PhysRevC.104.014601}.
Moreover, the neutron-energy resolution is better than that of the reactor due to the benefit of the pulsed neutron source.
In this paper, we report the angular distribution of $\gamma$ rays from the 1.33~eV p-wave resonance of $^{117}{\rm Sn}+n$ to the ground state of $^{118}{\rm Sn}$ at more setting angles by using a germanium detector assembly and the intense neutron beam at J-PARC.
The more detailed information of the angular distribution will progress the understanding of the $sp$-mixing model.

%% file: sections/section02.tex
\section{Experiment} \label{sec:experiment}

The measurement was performed using ANNRI installed at the beamline 04 (BL04) of the Material and Life Science Experiment Facility (MLF) in J-PARC.
A pulsed neutron beam was produced by the nuclear spallation reactions and was slowed down by a liquid hydrogen moderator.
The neutron beam was transported to the target position located at 21.5~m from the moderator surface, and a germanium detector assembly which consists of 22 germanium detectors was used to detect $\gamma$ rays emitted from the target. 
A more detailed explanation for the experimental setup is described in Ref.~\cite{PhysRevC.97.034662}.

The target was a tin plate with dimensions of $40 \times 40 \times 4~{\rm mm^3}$ and natural isotopic abundance, hence 7.68\% of $^{117}{\rm Sn}$. 
The chemical purity was 99.9\%. 
The total measurement time was about 65~hours with a proton beam power of 150~kW.
The deposited $\gamma$-ray energy calibration was performed by full-absorption peaks of $\gamma$ rays emitted from the $^{27}{\rm Al}(n,\gamma)$ reactions. 

\begin{figure}[tbp]
    \centering
    \includegraphics[width=.45\textwidth,angle=0]{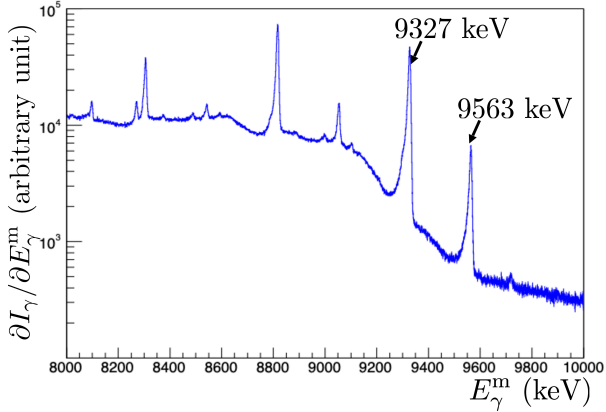}
    \caption{\label{fig:gammahist} Histogram of $\partial I_\gamma / \partial E^{\rm m}_\gamma$ in the range of 8-10~MeV.}
\end{figure}

Figure~\ref{fig:gammahist} shows the spectrum of the deposited $\gamma$-ray energy ($E^{\rm m}_\gamma$) in all detectors defined as $\partial I_\gamma / \partial E^{\rm m}_\gamma$. Here $I_\gamma$ denotes the number of detected $\gamma$ rays in the experiment.
The $\gamma$-transition from the compound state of $^{118}{\rm Sn}$ to the ground state with the energy of 9327~keV can be seen clearly.
It is known that the compound state in the p-wave resonance decays to the ground state of $^{118}{\rm Sn}$ directly.
Therefore, we focused our analysis on the peak with 9327~keV.
The $\gamma$-ray peak of 9563~keV stemmed from the $^{115}{\rm Sn}(n,\gamma)$ reactions.

%Figure: gamma-ray spectrum
%Figure: gamma-ray spectrum

\begin{figure}[tbp]
\centering
    \includegraphics[width=.45\textwidth,angle=0]{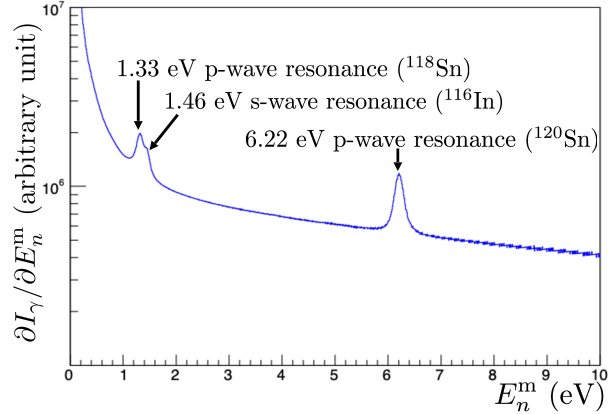}
\caption{\label{fig:tofhist} Histogram of $\partial I_\gamma / \partial E^{\rm m}_n$ in the range of 0-10~eV. In this histogram, $\gamma$ rays whose energies were below about 200~keV were eliminated. The peak of 1.33~eV is a p-wave resonance of $^{118}{\rm Sn}$.}
\end{figure}

The incident neutron energy $E_n^{\rm m}$ was calculated using the kinetic energy formula of classical mechanics substituting the time ($t^{\rm m}$) from the proton-beam injection into the MLF moderator to the arrival of neutrons at the tin target.
Figure~\ref{fig:tofhist} shows a histogram defined as $\partial I_\gamma / \partial E^{\rm m}_n$.
In this histogram, $\gamma$ rays whose energies were below about 200~keV were eliminated by energy thresholds of a data acquisition system.
The peak of 1.33~eV is a p-wave resonance of $^{118}{\rm Sn}$ of interest to us. 
One can see an s-wave resonance of $^{116}{\rm In}$ at 1.45~eV and a p-wave resonance of $^{120}{\rm Sn}$ at 6.22~eV.
Because the cross section of the 1.45~eV s-wave resonance is about $10^4$ times larger than that of the 1.33~eV p-wave resonance, their count rates become comparable even if the target is contaminated by 0.01\% of $^{115}{\rm In}$.
The pulse heights information was lost about 2\% of total $\gamma$-ray counts in the 1.33~eV p-wave resonance region, which was corrected in the following analysis.
The ratio of events that two signals are regarded as a single event was estimated as 0.3\% of total $\gamma$-ray counts in the 1.33~eV p-wave resonance region, which is negligibly small compared with the statistical error of the $\gamma$-ray peak counts focused in this study.

%Figure: neutron-energy spectrum
%Figure: neutron-energy spectrum

The cross section for the $(n,\gamma)$ reaction is expressed by the Breit-Wigner function described by resonance parameters: resonance energy $E_r$, total angular momentum $J_r$, orbital angular momentum $l_r$, $\gamma$ width $\Gamma^\gamma_r$, statistical factor $g_r$, and neutron width $\Gamma^n_r$. Here the subscript $r$ means an s-wave or a p-wave resonances.
%The resonance parameters of the p-wave resonance and a negative s-wave resonance were measured using a $^{117}$Sn-enriched tin target.
The resonance parameters of the p-wave resonance was measured using a $^{117}$Sn-enriched tin target.
The measurement time was about 3~hours with the proton beam power of 525~kW.
The resonance parameters were determined by fitting the p-wave resonance gated with $E^{\rm m}_\gamma \ge 2~{\rm MeV}$ using the Breit-Wigner function convoluted with the Doppler broadening effect and the time structure of the pulsed neutron beam~\cite{NIMA.736.66}.
In this work, the resonance energy and the $\gamma$ width of the p-wave resonance were obtained as $E_{\rm p} = 1.331 \pm 0.002~{\rm eV}$ and $\Gamma^\gamma_{\rm p} = 133 \pm 5~{\rm meV}$. 
In the fitting, other resonance parameters were fixed at published values: $J_{\rm p} = 1$, $l_{\rm p} = 1$, and $g_{\rm p}\Gamma^n_{\rm p} = 1.38 \times 10^{-4}~{\rm meV}$~\cite{PhysRevC.59.2836}.

%Table~\ref{tab:ResonanceParameter} summarizes the resonance parameters used in this work.
\begin{comment}
\begin{table}[H]
    \centering
    \caption{\label{tab:ResonanceParameter}
    Resonance parameters of the neutron resonances of $^{117}{\rm Sn}+n$. 
    The values with index (a) and (b) are taken from Refs.~\cite{PhysRevC.59.2836} and \cite{Mughabghab}, respectively.
    Other values were determined by fitting.}
    \begin{tabular}{c||c|c|c|c|c}
        $r$ & $E_r$ [eV] & $J_r$ & $l_r$ & $\Gamma^\gamma_r$ [meV] & $g_r \Gamma^n_r$ [meV] \\
         \hline
        s & -29.2$^{(a)}$ & 1$^{(b)}$ & 0 & $77.0 \pm 2.0$ & 29.9$^{(a)}$\\
         p & $1.331 \pm 0.002$ & 1 & 1 & $133 \pm 5$ & $1.38 \times 10^{-4}$~$^{(a)}$
    \end{tabular}
\end{table}
\end{comment}

%% file: sections/section03.tex
\section{Analysis} \label{sec:analysis}

\begin{figure}[tbp]
    \centering
    \includegraphics[width=.45\textwidth,angle=0]{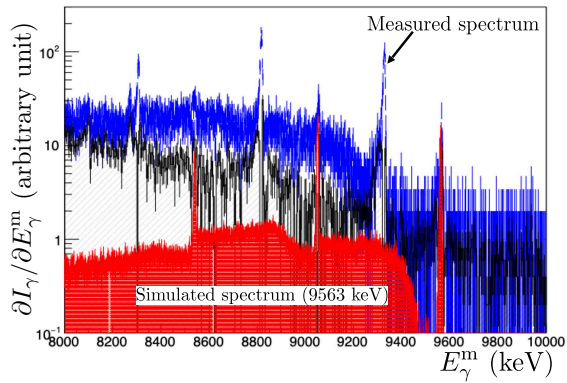}
    \caption{\label{fig:BGgammahist} Histograms of $\partial I_\gamma / \partial E^{\rm m}_\gamma$ for each background event. Blue-open histogram is the measured spectrum. Red-hatched one is a simulated spectrum for monochromatic energy of 9563~keV. This was scaled so that the intensity of the full-absorption peak matched that of the measured one. Black-striped histogram is the spectrum of pileup events which were obtained by subtracting the red-hatched spectrum and a simulated spectrum of monochromatic energy of 9327~keV.}
\end{figure}

%\subsection{Corrections} \label{subsec:corrections}
%\subsubsection{Subtraction of background events} \label{subsubsec:BGsubtraction}
The histogram gated with the 9327~keV full-absorption peak includes background events from other sources than the resonance of interest in $^{118}{\rm Sn}$.
Here, the gate range was defined as the full width at quarter maximum (FWQM).
There are two main kinds of background events.
One is the Compton scattering of $\gamma$ rays with 9563~keV. 
The other is caused by pileup events due to simultaneous detection of multi $\gamma$ rays.
These background events must be subtracted.
First, the number of such events in the signal region was estimated using a GEANT4 simulation~\cite{JINST.13.P02018}.
In this simulation, a response function of the detector assembly can be calculated, and a spectrum of monochromatic $\gamma$-ray energy can be reproduced by emitting $\gamma$ rays from the target position.
Figure~\ref{fig:BGgammahist} shows the histograms of $\partial I_\gamma / \partial E^{\rm m}_\gamma$ for each background event.
Histograms gated with the background regions (9563~keV peak and energies higher than 9600~keV) were scaled so that number of events in the signal region matched that of simulated ones in the $\gamma$-ray energy spectrum. 
After that, they were subtracted from the histogram gated with the signal regions.

%Figure: B.G. gamma-ray spectrum
%Figure: B.G. gamma-ray spectrum

%\subsubsection{Normalization of beam intensity} \label{subsubsec:BeamIntensity}
In the epithermal energy region, the neutron-beam intensity at J-PARC increases for lower neutron energies as a result of moderation.
The histogram defined by $\partial I_\gamma / \partial E^{\rm m}_n$ must be normalized to obtain the neutron-energy dependence of the 1.33~eV p-wave resonance shape.
The energy dependence of incident neutron beam was obtained by measuring the 477.6~keV $\gamma$ rays emitted by the $^{10}{\rm B}(n,\alpha \gamma)^{7}{\rm Li}$ reactions with a boron target.
The $^{10}{\rm B}(n,\alpha \gamma)^{7}{\rm Li}$ reaction is suitable to obtain the energy dependence of the neutron beam because there are no resonances in the epithermal region ($\mathcal{O}$(1~eV)).
The beam intensity $\psi(E_n^{\rm m})$, as a function of neutron energy $E_n$, can be represented as
\begin{equation}
\psi(E_n^{\rm m}) = \frac{I(E_n^{\rm m})}{\sigma(E_n) \epsilon T},
\end{equation}
where $\sigma(E_n)$ is the cross section of the $^{10}{\rm B}(n,\alpha \gamma)^{7}{\rm Li}$ reaction, which is $3837\pm 9$~barn for 2200~m/s neutrons~\cite{Mughabghab} and is assumed to be dependent on neutron velocity $v$ based on the $1/v$ law. The $I(E_n^{\rm m})$ is the number of the 477.6~keV $\gamma$ rays detected during the measurement time $T$, and $\epsilon$ is the detection efficiency of each germanium detector at 477.6~keV.

The effective photopeak efficiency of each germanium detector, including the solid angle coverage of each crystal, is different each other. 
The effective photopeak efficiency was determined relatively based on the assumption that prompt $\gamma$ rays from $^{14}$N(n,$\gamma$) reactions of a melamine target were emitted isotropically.

%\subsection{Angular distribution of p-wave resonance} \label{subsec:AngularDistribution}
%Figure: p-wave resonances in each angle
\begin{figure*}[tbp]
    \centering
    \includegraphics[width=.85\textwidth]{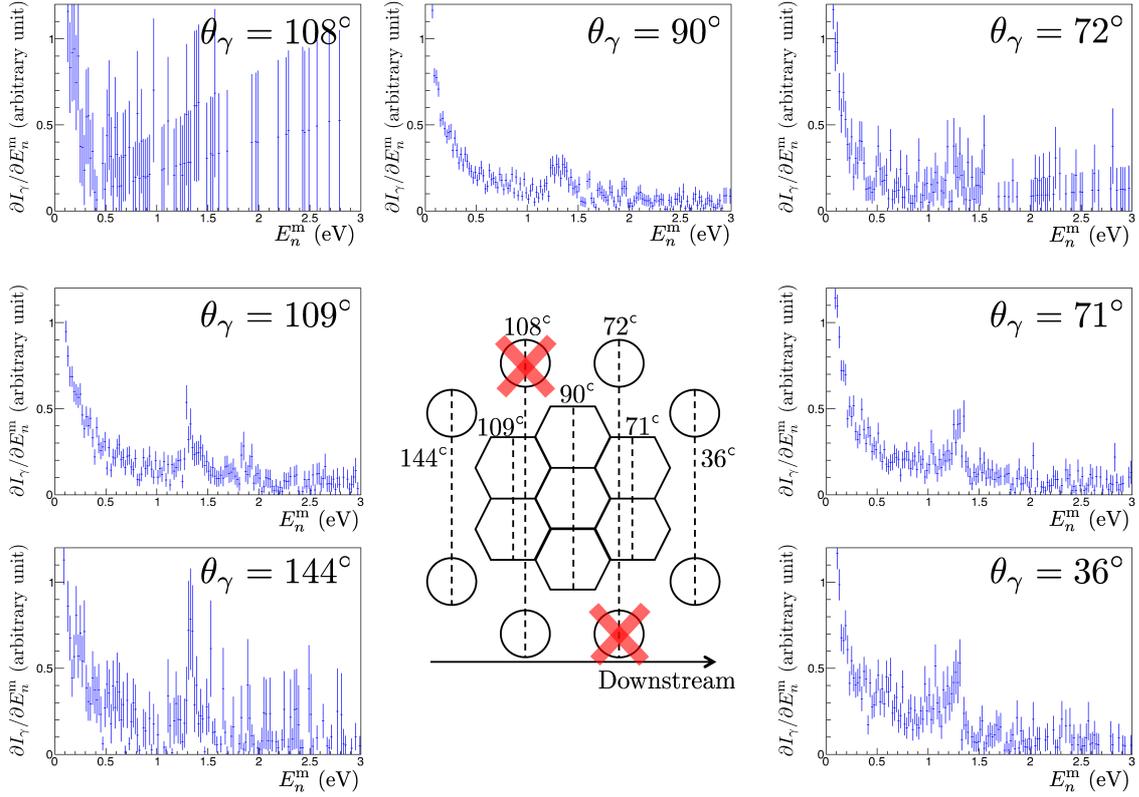}
    \caption{\label{fig:AngularDistribution} Histograms of $\partial I_\gamma / \partial E^{\rm m}_n$ around the 1.33~eV p-wave resonance in $^{118}{\rm Sn}$ for various angles accessible at ANNRI. The central figure shows the placement and shape of each germanium crystals. The detectors marked by red crosses were not used in this measurement.}
\end{figure*}
%Figure: p-wave resonances in each angle
Figure~\ref{fig:AngularDistribution} shows histograms as a function of $E^{\rm m}_n$ around the 1.33~eV p-wave resonance for various angles.
The $E^{\rm m}_n$ dependence of the angular distribution can be observed as an asymmetric resonance shape in the histograms of $\partial I_\gamma / \partial E^{\rm m}_n$.

%% file: sections/section04.tex
\section{Discussion} \label{sec:discussion}

To evaluate the $E^{\rm m}_n$ dependence of the angular distribution of $\gamma$ rays, an asymmetry $A_{\rm LH}(\theta_\gamma)$ were used as in the analysis for lanthunum~\cite{PhysRevC.97.034662}.
The asymmetry $A_{\rm LH}(\theta_\gamma)$ is defined as
\begin{align}
    A_{\rm LH}(\theta_\gamma) = \frac{N_{\rm L}(\theta_\gamma) - N_{\rm H}(\theta_\gamma)}{N_{\rm L}(\theta_\gamma) + N_{\rm H}(\theta_\gamma)},
\end{align}
where $\theta_\gamma$ is the $\gamma$-ray emission angle with respect to the neutron momentum,
and this asymmetry can be parameterized as 
\begin{align}
    A_{\rm LH}(\theta_\gamma) = A\cos\theta_\gamma + B.
    \label{eq:AsymmetryParameter}
\end{align}
Here, the subscripts L and H represent "low energy region" and "high energy region", respectively.
The regions for integration were defined using the p-wave resonance energy $E_{\rm p}$ and the total resonance width $\Gamma_{\rm p}$ as follows;
$E_{\rm p}-2\Gamma_{\rm p} < E_n < E_{\rm p}$ for $N_{\rm L}(\theta)$ and $E_{\rm p} < E_n < E_{\rm p}+2\Gamma_{\rm p}$ for $N_{\rm H}(\theta)$.
The total resonance width is defined as $\Gamma_{\rm p} = \Gamma^\gamma_{\rm p} + \Gamma^n_{\rm p}$.
Here, the resonance parameters, $E_{\rm p}$, $\Gamma^\gamma_{\rm p}$, and $\Gamma^n_{\rm p}$, were the values described in Sec.~\ref{sec:experiment}.
%Here, the resonance parameters, $E_{\rm p}$, $\Gamma^\gamma_{\rm p}$, and $\Gamma^n_{\rm p}$, were the values described in Table~\ref{tab:ResonanceParameter}.

\begin{figure}[htbp]
    \centering
    \includegraphics[width=.45\textwidth]{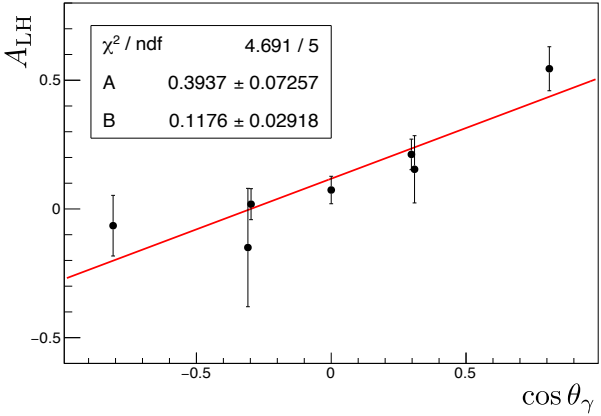}
    \caption{Fitting result for $A_{\rm LH}(\theta_\gamma)$ using Eq.~(\ref{eq:AsymmetryParameter}). The angular-dependent parameter $A$ were obtained to be a significant value.}
    \label{fig:Asym_fixedfit}
\end{figure}

The asymmetry $A_{\rm LH}(\theta_\gamma)$ can be fitted using Eq.~(\ref{eq:AsymmetryParameter}), and the fitting result is shown in Fig.~\ref{fig:Asym_fixedfit}.
The parameters of the $A_{\rm LH}(\theta_\gamma)$ were obtained as
\begin{align}
A = 0.394 \pm 0.073 ~{\rm and}~ B = 0.118 \pm 0.029.
\end{align}
Significant $E^{\rm m}_n$ dependence of the angular distribution was observed in the vicinity of the p-wave resonance.
%Future measurements of other angular-dependent terms with polarized and unpolarized neutron beam may verify the $sp$-mixing model of compound states.

%% file: sections/section05.tex
\section{Conclusion} \label{sec:conclusion}

The experiment was conducted for measuring the $E^{\rm m}_n$ dependence of angular distribution of $\gamma$ rays in the transition from the p-wave resonance of $^{117}{\rm Sn}+n$ to the ground state of $^{118}{\rm Sn}$ at BL04 ANNRI in J-PARC MLF.
The $E^{\rm m}_n$ dependence of angular distribution of $\gamma$ rays was significantly observed, and the angular-dependent asymmetric resonance shape was evaluated using the asymmetry value $A_{\rm LH}(\theta_\gamma)$.
The angular dependence of the asymmetry $A_{\rm LH}(\theta_\gamma)$ was obtained as $A_{\rm LH}(\theta_\gamma) = A\cos \theta_\gamma +B$ with $A = 0.394\pm 0.073$ and $B = 0.118 \pm 0.029$.
This result and future measurements of angular distributions of $\gamma$ rays with polarized neutron beam will allow for cross checking with the previous result, which will progress the understanding of $sp$-mixing model of compound states.

%% file: sections/acknowledgements.tex
\section*{Acknowledgements} \label{sec:acknowledgements}
%    \lipsum[8]
The authors would like to thank the staff of ANNRI for the maintenance of the detectors, and MLF and J-PARC for operating the accelerator and the neutron-production target.
The experiments at MLF of J-PARC were performed under the program (Proposals No. 2017L2000, 2017A0158, 2018B0193, and 2019A0184).
This work was supported by the Neutron Science Division of KEK as an S-type research project with program number 2018S12.
This work was partially supported by MEXT KAKENHI Grant No. JP19GS0210 and JSPS KAKENHI Grant No. JP17H02889.